# Torque-operated gradient-index axicon

**Yurij Vasylkiv, Mykola Smyk, Ihor Skab and Rostyslav Vlokh ***

*Institute of Physical Optics, 23 Dragomanov Street, 79005 Lviv, Ukraine*

*\* Correspondence author: vlokh@ifo.lviv.ua*

**Abstract**

We describe a gradient-index axicon based on twisting of crystals. We demonstrate that the focal length of the axicon can be efficiently operated by the torsion moment. The working analytical relations describing the focal length of the axicon and its dependence on different geometrical parameters as well as the torsion moment has been derived.

## Introduction

An axicon lens that transforms Gaussian beams into non-diffractive Bessel beams [1, 2] has been suggested by J. H. McLeod in 1954 [3, 4]. In fact, one cannot produce ideal Bessel beams with axicons since a uniform plane wave is not realizable. However, using the collimated incident Gaussian beams, one can generate so-called Bessel–Gaussian beams, which manifest greatly reduced diffraction. Such beams are efficiently used for optical trapping, high-precision drilling of holes, in optical coherence tomography, etc. [5–8]. Moreover, high-order Bessel beams carrying a nonzero orbital angular momentum can be applied for transmitting, coding and decoding information [9].

The axicons are often fabricated in the shape of a conical lens formed by a plane and conical surfaces. Recently gradient-index axicons have been suggested [10, 11]. Instead of a conical shape, they are based on a specific spatial distribution of refractive index created inside a parallel plate, using various technologies. It is obvious that fabrication of axicons with the conical shape and long enough focal length is quite difficult because the angle at cone apex should be close to $p$, while the cone vertex should be produced with high accuracy in order to avoid the appearance of distortions of the phase front. On the contrary, the gradient-index axicons with flat parallel surfaces and a desired radial dependence of the refractive index can be produced, using well-known techniques such as ionic exchange or chemical vapor deposition.

Notice that both of the conical axicon and the gradient-index axicon represent passive optical elements, which do not permit operating by their characteristics. Nonetheless, recently an adjustable axicon mirror has been proposed in order to produce the Bessel beams [12], of which characteristics are operated while deforming a silicon wafer in its center. Such an optical element is cheap enough and can be useful in cold-atom manipulation. However, the device can work only as a reflecting axicon. Furthermore, a tunable acoustic gradient-index lens has been suggested for generating the Bessel beams [13]. This axicon lens operates basing on acoustooptic diffraction of the incident Gaussian beam on the acoustic waves generated by a circular piezoelectric transducer. The device reveals high operation speed and is operated by changing acoustic-wave frequency. Variations in the driving frequency and the amplitude of the acoustic wave propagating in the lens filled by a liquid are equivalent to changing the axicon cone angle. However, the frequency band of the piezoelectric transducer is usually limited by the thickness of the latter and so cannot be made wide enough. Notice that the main principles used for designing the flat gradient-index axicons are (i) creation of a desired radial dependence of the refractive index and (ii) elaboration of physical principles for operating the slope of this dependence. In principle, the both conditions can be satisfied issuing from the generally known parametric optical effects, i.e. electrooptics or piezooptics induced by inhomogeneous external fields.

The present work is devoted to the studies of possibilities for operating the main characteristics of the flat gradient-index axicons. In particular, we will study an adjustable flat parallel axicon lens formed by torsion of crystals.

## The principles of operation of a gradient-index torsion axicon

It is known [14] that, under torsion of a cylindrical rod around its geometrical axis, there appears a shear stress which may be presented by the relation

$$s_m = \frac{2M_Z}{pR^4}(Xd_{4m} - Yd_{5m}).  \quad (1)$$

Here $Z$ axis coincides with the torsion axis, $X$ and $Y$ are orthogonal to $Z$, $M_Z = \int_S (r \times P) dS$ is the torque, $d_{nm}$ the Kronecker delta, $R$ the cylinder radius, $S$ the square of the cylinder basis, and $P$ the mechanical load. As a result, we have the two shear components of the stress tensor, $s_4$ and $s_5$. The piezooptic effect is described by changes in the optical impermeability coefficients $\Delta B_i$ (or the refractive indices $n$, since $B_i = (1/n^2)_i$) imposed by the action of mechanical stresses $s_m$ (see, e.g., [15]):

$$\Delta B_i = B_i - B_i^0 = p_{im} s_m, \quad (2)$$

where $p_{im}$ is a fourth-rank piezooptic tensor, and $B_i$ and $B_i^0$ the impermeability tensors of stressed and free samples, respectively. As shown in our works [16–18], application of torsion stresses leads to specific spatial distributions of the refractive indices, optical birefringence, and optical indicatrix rotation. In particular, this occurs in the crystals belonging to the cubic or trigonal systems [19, 20]. An appropriate example is LiNbO$_3$ crystals under conditions that a laser beam propagates along the optic axis and the torsion is applied around the same direction [21].

What is the most important, a special singular point of zero birefringence belonging to the torsion axis has been revealed in the geometrical center of $XY$ cross section of a sample, which corresponds to zero shear stress components $s_5$ and $s_4$. The birefringence has been found to increase linearly with increasing distance from that geometrical center and so the birefringence distribution has a conical shape in the coordinates $X, Y, \Delta n$ [21]. The spatial distribution of the optical indicatrix rotation angles $z_Z$ in the crystals belonging to the point symmetry group 3m (the case of lithium niobate) is defined by the relation $z_Z = j/2$, where $j$ is the polar angle ($X = r\cos j$ and $Y = r\sin j$). In fact, we have found experimentally [21] that an optical vortex appears in the optical system consisting of a right-handed circular polarizer, a LiNbO$_3$ crystal twisted around the Z axis, and a left-handed circular polarizer. The change in the phase of outgoing light coincides with the tracing angle, and the vortex induced by the torsion in our case has a topological charge equal to unity. It is obvious that the sign of the vortex charge would depend on the 'sign' of the incident circularly polarized wave. However, in our case the original cause is inhomogeneous spatial distribution of the birefringence.

While analyzing the properties of a lens of torsion type, one should be interested in the radial dependence of the refractive indices in, e.g., the crystals of the point symmetry group 3m mentioned above. The dependences of the refractive indices on the $X$ and $Y$ coordinates are given by the relations [17]

$$n_1 = n_o - \frac{n_o^3}{2} p_{14} \sqrt{s_4^2 + s_5^2} = n_o - n_o^3 p_{14} \frac{M_Z}{pR^4} \sqrt{X^2 + Y^2} = n_o - n_o^3 p_{14} \frac{M_Z}{pR^4} r, \quad (3)$$

$$n_2 = n_o + \frac{n_o^3}{2} p_{14} \sqrt{s_4^2 + s_5^2} = n_o + n_o^3 p_{14} \frac{M_Z}{pR^4} \sqrt{X^2 + Y^2} = n_o + n_o^3 p_{14} \frac{M_Z}{pR^4} r. \quad (4)$$

It is seen that the refractive indices $n_1$ and $n_2$ depend linearly on the polar coordinate $r$ and the torque $M_Z$. The torsion-induced birefringence forms a conical surface in the $X$ and $Y$ coordinates, with the singular zero-value point at the origin of the coordinate system:

$$\Delta n_{12} = -n_o^3 p_{14} \sqrt{s_4^2 + s_5^2} = -2n_o^3 p_{14} \frac{M_Z}{pR^4} \sqrt{X^2 + Y^2}. \quad (5)$$

Using our earlier data for the piezooptic coefficient $p_{14} = (8.87 \pm 0.28) \times 10^{-13}\,\mathrm{m^2/N}$ of LiNbO$_3$ [22] and the data for the refractive index ($n_o = 2.28647$ for $\lambda = 632.8\,\mathrm{nm}$ [23]), one can calculate the coordinate dependences of the refractive indices for these crystals under different torques (see Fig. 1).

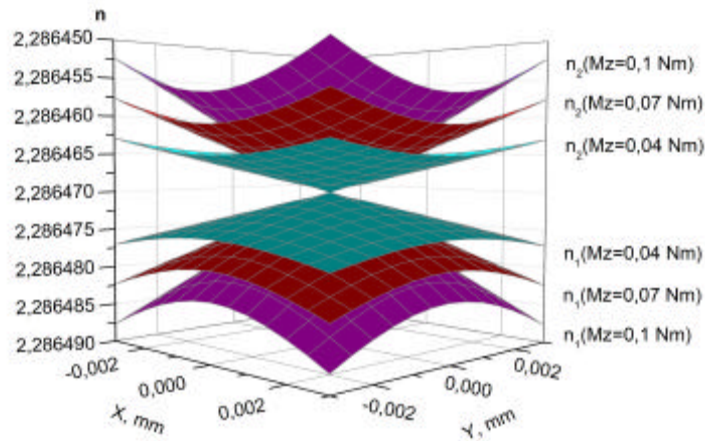

(a)

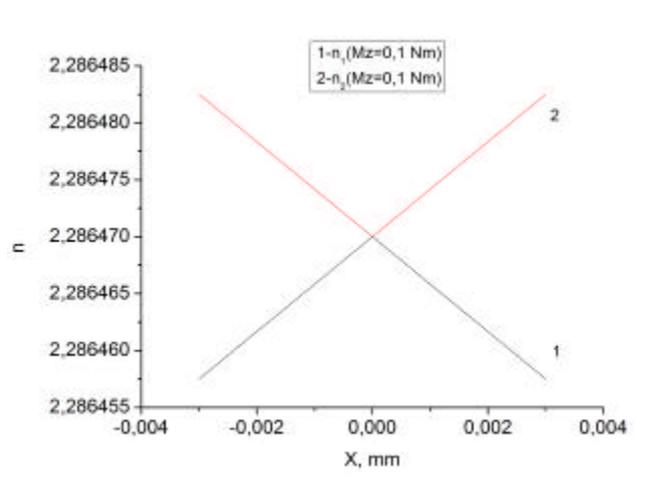

(b)

Fig. 1. Calculated *XY* coordinate dependences of the refractive indices for LiNbO$_3$ crystals under different torsion moments applied: $M_Z = 0.04;\ 0.07;\ 0.1\,\mathrm{N \times m}$ (a); cross section of the refractive indices surface at $M_Z = 0.1\,\mathrm{N \times m}$ by the *Y*=0 plane (b) ($\lambda = 632.8\,\mathrm{nm}$).

As follows from Fig. 1, the refractive indices acquire the change of $\sim 1.5 \times 10^{-5}$ under the torque of $\sim 0.1\, N \times m$. Moreover, the sign of the refractive index changes depends on that of the torsion moment. From the equation for the cross section of the optical indicatrix,

$$(B_1 + p_{14} s_4(X))X^2 + (B_1 - p_{14} s_4(X))Y^2 + 2 p_{14} s_5(Y)XY = 1, \qquad (6)$$

it follows that the optical indicatrix rotates by 180 deg around the geometrical center of the $XY$ cross section when the angle $j$ is changed by 360 deg (Fig. 2). In other words, the angle of the optical indicatrix rotation is given by

$$\tan 2z_Z = \frac{s_{13}}{s_{23}} = \frac{Y}{X} = \tan j \quad \text{or} \quad z_Z = j/2. \qquad (7)$$

Notice that the condition $X \perp m$ holds true for our coordinate system, i.e. the $X$ axis is perpendicular to one of the mirror symmetry planes.

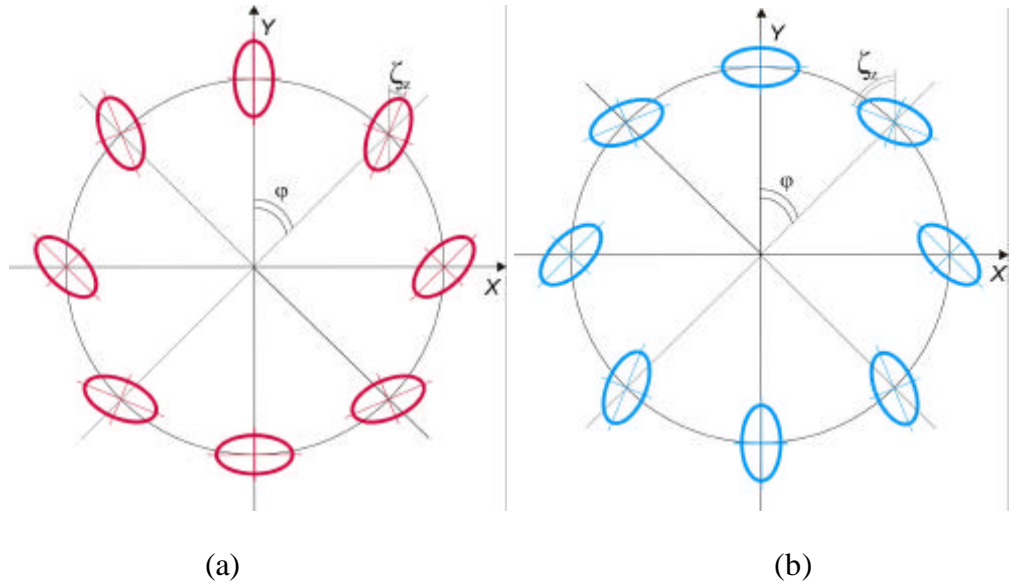

(a) (b)

Fig. 2. Distributions of the optical indicatix rotation in LiNbO$_3$ crystals for the two opposite torsion moments, as follows from Eqs. (3), (4) and (7). Orientations of cross sections of the optical indicatrix are introduced by ellipses.

It follows from the analyses presented above that the crystals subjected to torsion can act as lenses. However, polarization of the incident light beam should be linear and spatially distributed around the beam axis, being rotated by the angle $z_Z$. Moreover, the directions of the elec-tric field oscillations for the incident optical wave should coincide with the eigenvectors of the optical impermeability tensor (see Fig. 3). A twisted crystalline rod will act as a converging or diverging lens in the tow alternative cases of mutually orthogonal directions of the incident light polarization. In addition, the sign of the lens can be changed from positive to negative by changing the torque sign.

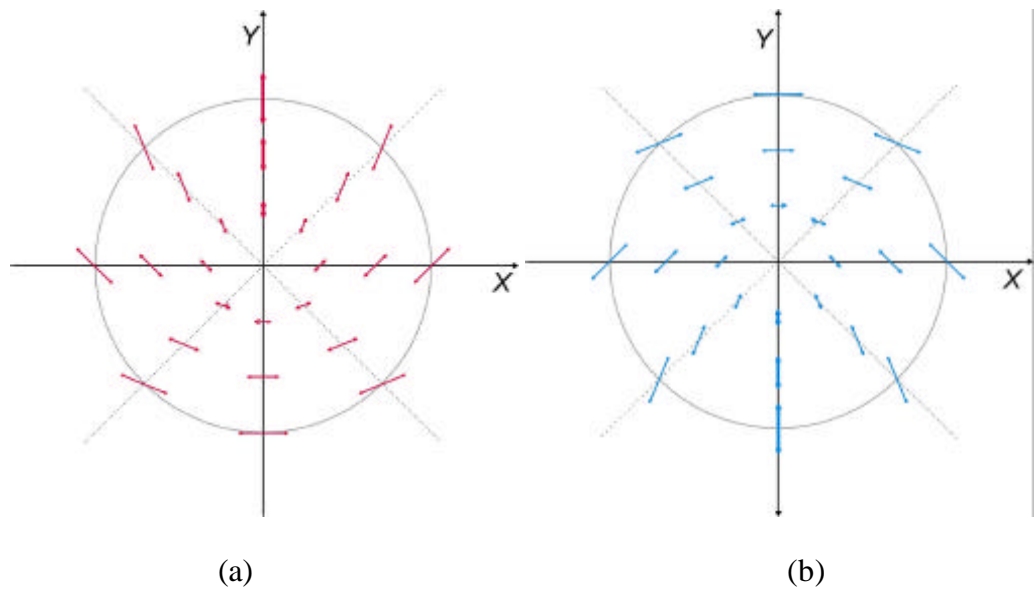

(a) (b)

Fig. 3. Spatial distributions of directions of the electric field oscillations for the incident optical wave presented for the case of converging (a) and diverging (b) lenses, with respect to the cross sections of the optical indicatrix shown in Fig. 2,a.

It is known [24, 25] that ordinary gradient-index lenses are usually characterized by a polynomial or even-power radial dependence of the refractive index, in order to focus a point light source into a point. At the same time, axicons should transform a point source into a focal line. It is a presence of odd-power terms in the dependence $n(r)$ or even a linear term that turns, e.g., a biconvex lens into an axicon or pseudoaxicon [24–26]. In our further calculations we will follow the approach presented in the work [24]. When the gradient of the refractive index is small, one can consider a twisted crystalline rod as a thin lens. Then the transverse beam displacement during its propagation inside the rod can be considered as infinitely small and so can be neglected. A scheme of the wave front change under propagation of a collimated light beam through a twisted crystalline rod is presented in Fig. 4.

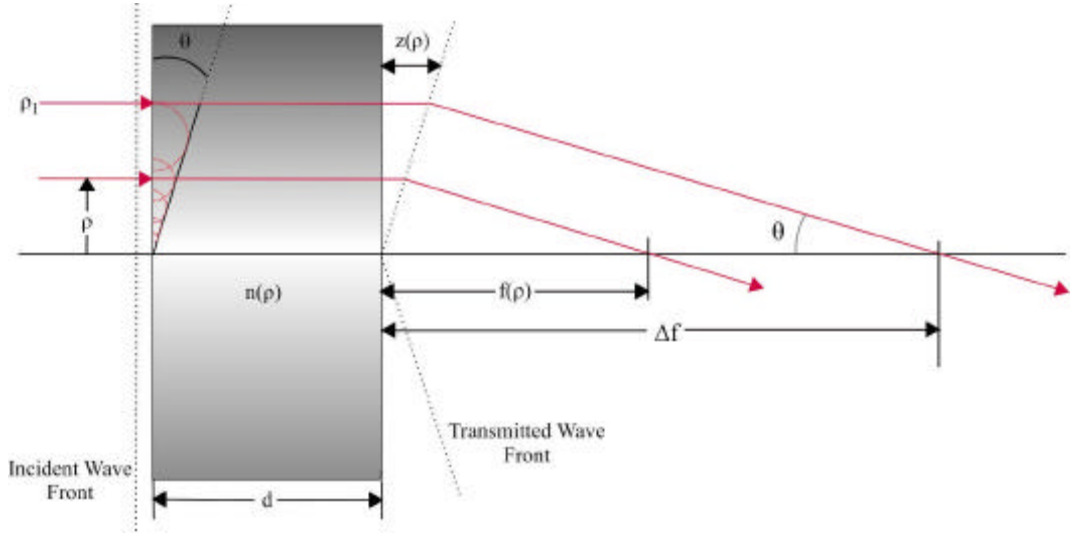

Fig. 4. A scheme of light wave front changes occurring when a collimated beam propagates through a twisted crystalline rod.

It is easy to show that the focal length in our case is given by the relation

$$f(\mathbf{r}) = \frac{\mathbf{r}}{r_1}\Delta f, \qquad (8)$$

where $\mathbf{r}$ is the radial position of an incident ray at the front surface, $r_1$ the maximum height of the ray, and $\Delta f$ is equal to the focal length at $\mathbf{r} = r_1$. Because the geometric angle $\mathbf{q}$ of the ray is equal to the slope of the wave-front normal, the wave front is described by

$$\tan \mathbf{q} = -\frac{dz}{d\mathbf{r}} \approx -\frac{\mathbf{r}}{f(\mathbf{r})} \quad (\text{at } z(\mathbf{r}) \ll f(\mathbf{r})), \qquad (9)$$

where $z(\mathbf{r})$ is the wave front displacement. Combining Eq. (9) with Eq. (8), we get

$$z(\mathbf{r}) = \int dz = \frac{r_1}{\Delta f}\int d\mathbf{r} = \frac{r_1}{\Delta f}\mathbf{r}. \qquad (10)$$

Accounting for a small size of the gradient, one can express the optical path as

$$n_o d = n(\mathbf{r})d + z(\mathbf{r}), \qquad (11)$$

while the dependence of the refractive index on the coordinate is determined by the formula

$$n(\mathbf{r}) = n_o - z(\mathbf{r})/d, \qquad (12)$$

with $d$ being the thickness of the crystalline rod. Taking Eqs. (10) and (12) into consideration, we write the coordinate dependence of the refractive index in the following form:

$$n(\mathbf{r}) = n_o - \frac{r_1}{d\Delta f}\mathbf{r}. \qquad (13)$$

Let us notice that Eqs. (3) and (4), on the one side, and Eq. (13), on the other are of the same structure. Then the focal length is determined by the relation

$$\Delta f = \frac{r_1 p R^4}{d n_o^3 p_{14} M_Z}. \qquad (14)$$

Calculating the focal distance for the case of LiNbO$_3$ crystalline rod with the geometrical parameters $R = 3$ mm, $r_1 = 2$ mm, $d = 13$ mm and $M_Z = 0.1$ N×m, one obtains $\Delta f$ equal to 36.9 m. Dependences of the focal length on the torque for different radiuses of the incident beam are presented in Fig. 5.

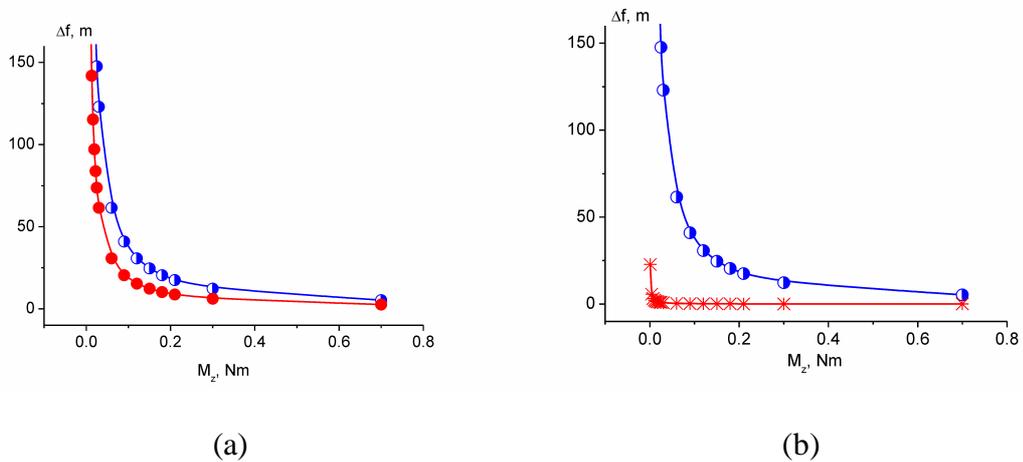

(a)          (b)

Fig. 5. Dependences of focal length on the torsion moment for a LiNbO$_3$ crystalline rod: (a) the geometrical parameters are $R = 3$ mm and $d = 13$ mm, and the radiuses of the incident beam are equal to $r_1 = 1$ mm (semi-full circles) and $r_1 = 2$ mm (full circles); (b) the same for the cases of $R = 3$ mm, $d = 13$ mm and $r_1 = 2$ mm (semi-full circles) and $R = 1$ mm, $d = 13$ mm and $r_1 = 1$ mm (stars). All the calculations are performed for the light wavelength $l = 632.8$ nm.

Since the focal length is inversely proportional to the torsion moment, it decreases with increasing torque. Notice that the focal length can be changed in a wide enough range by varying the geometrical parameters. For instance, doubling the beam radius would result in the same change of the focal length (see Fig. 5,a). At the same time, reducing the rod radius by three times, with simultaneously decreasing the beam radius by two times (then the radiuses of the rod and the beam being equal), would provide a notable decrease in the $\Delta f$ parameter as (see Fig. 5,b).

## Conclusions

In the present work we suggest a gradient-index axicon based on torsion of crystals. In particular, we derive the working analytical relations describing the focal length of the axicon and its dependence on different geometrical parameters as well as the torsion moment. It is shown that the focal length of the axicon can be operated by the torque in a wide enough range. Functionality of the axicon fabricated from $LiNbO_3$ crystals is demonstrated using calculations of the focal length as a function of the torque magnitude. Large focal lengths easily achievable with our technique, which can exceed a hundred of meters, can be used in such branches as telecommunications or transmitting optical beams in through interstellar medium.

In fact, generation of the first-order Bessel beam bearing a singly charged optical vortex has been demonstrated in our recent work [21] for a twisted $LiNbO_3$ crystal placed in between the crossed circular polarizers. Experimental verifications of the generation of zero-order Bessel beam using our axicon can be associated with some potential difficulties, namely a need in formation of specifically polarized incident beams (see Fig. 3). Nonetheless, such a hollow beam can be created, e.g., while using a conical refraction effect in optically biaxial crystals [27–29] or a liquid crystalline matrix with an embedded topological defect of the strength ½ [30]. Experimental studies of the torsion-operated gradient-index axicon will be a topic of our forthcoming paper.